\documentclass[aps,prl,twocolumn,showpacs,groupedaddress]{revtex4-1}

\usepackage{epsfig}
\usepackage{graphicx}
\usepackage{dcolumn}
\usepackage{bm}
\usepackage[colorlinks=true,dvipdfm]{hyperref}
\usepackage{amssymb}

\begin{document}
\title{Compelling evidence for the theory of dynamic scaling in first-order phase transitions}

\author{Fan Zhong}
\affiliation{State Key Laboratory of Optoelectronic Materials and
Technologies and School of Physics, Sun Yat-sen
University, Guangzhou 510275, People's Republic of China}

\date{\today}

\begin{abstract}
Matter exhibits phases and their transitions. These transitions are classified as first-order phase transitions (FOPTs) and continuous ones. While the latter has a well-established theory of the renormalization group, the former is only qualitatively accounted for by classical theories of nucleation, since their predictions often disagree with experiments by orders of magnitude. A theory to integrate FOPTs into the framework of the renormalization-group theory has been proposed but seems to contradict with extant wisdom. Here we show first that classical nucleation and growth theories alone cannot explain the FOPTs of the paradigmatic two-dimensional Ising model driven by linearly varying an externally applied field. Then we offer compelling evidence that the transitions agree well with the renormalization-group theory when logarithmic corrections are properly considered. This unifies the theories for both classes of transitions and FOPTs can be studied using universality and scaling similar to their continuous counterpart.
\end{abstract}

\maketitle

Matter as a many-body system exists in various phases and/or their coexistence and its diversity comes from phase changes. It thus exhibits just phases and their transitions. These transitions are classified as first-order phase transitions (FOPTs) and continuous ones. Although the phases can be studied by a well-established framework and the continuous phase transitions have a well-established theory of the renormalization group (RG) that has predicted precise results in good agreement with experiments, the FOPTs gain a different status in statistical physics. They proceed through either nucleation and growth or spinodal decomposition~\cite{Gunton83,Bray,Binder2}. Although classical theories of nucleation~\cite{Becker,Becker1,Becker2,Zeldovich,books,books1,books2,Oxtoby92,Oxtoby921,Oxtoby922,Oxtoby923} and growth~\cite{Avrami,Avrami1,Avrami2} correctly account for the qualitative features of a transition, even an agreement in the nucleation rate of just several orders of magnitude between theoretical predictions and experimental and numerical results is considered as a feat~\cite{Oxtoby92,Oxtoby921,Oxtoby922,Oxtoby923,Filion,Filion1,Filion2}. A lot of improvements have thus been proposed and tested in the two-dimensional (2D) Ising model whose exact solution is available. One theory of nucleation, called FT hereafter, considers field theoretic corrections to the classical theory~\cite{Langer67,Gunther}. Its field dependence was quantitatively verified for a constant applied magnetic field $H$ that directs oppositely to the equilibrium magnetization $M_{\rm eq}$ at a temperature $T$ below the critical temperature $T_c$ by Monte Carlo simulations~\cite{Rikvold94}. By employing the results of such relaxation processes, FT was also shown to accurately produce numerical results of hysteresis loop areas in a single droplet (SD) regime in which only a single droplet nucleates and grows quickly throughout the system~\cite{Sides98}. So was in a multidroplet (MD) regime where many droplets nucleate and grow even in the case of a sinusoidally varying $H$ by using Avrami's growth law~\cite{Avrami,Ramos} and an adiabatic approximation~\cite{Sides99}. In this regime, an adjustable parameter was needed to match the area of just one frequency but then yielded good results for others~\cite{Sides99}. Another theory, referred to as BD below, adds appropriate corrections to the droplet free energy of Becker and D\"{o}ring's nucleation theory~\cite{Becker2}. Such a theory was found to accurately predict nucleation rates for the 2D Ising model without adjustable parameters~\cite{Ryu,Ryu1}.

However, it is well-known that classical nucleation theories are not applicable in spinodal decompositions in which the critical droplet for nucleation is of the lattice size and thus no nucleation is needed~\cite{Gunton83}. Although sharply defined spinodals that divide the two regimes of the apparently different dynamic mechanisms do not exist for systems with short-range interactions contrary to the mean-field case which has long-range interactions~\cite{Gunton83,Bray,Binder2}, it is generally believed that there exists a crossover region between them at least at the early stage of an FOPT for systems with short-range interactions~\cite{Gunton83,Bray,Binder2}. One may then characterize this crossover by fluctuation-shifted mean-field spinodals and expand near such instability points below $T_c$ of a usual $\phi^4$ theory that describes the critical behavior of the Ising model. This results in a $\phi^3$ theory for the FOPT due to the lack of the up--down symmetry in the expansion~\cite{Zhongl05,zhong16}. An RG theory for the FOPT can then be set up in parallel to that for the critical phenomena, giving rise to universality and dynamic scaling characterized by ``instability'' exponents corresponding to the critical ones. The primary qualitative difference is that the nontrivial fixed points of such a theory are imaginary in values and are thus usually considered to be unphysical, though the instability exponents are real. Yet, it is later shown that counter-intuitively imaginariness is physical in order for the $\phi^{3}$ theory to be mathematically convergent, since at the instability points, the unstable degrees of freedom of the system flows to the fixed points upon coarse graining~\cite{Zhonge12}. Moreover, the other degrees of freedom that need finite free energy costs for nucleation are coarse-grained away with the costs and are thus irrelevant to the transition~\cite{Zhonge12}. This indicates that nucleation is irrelevant to the scaling. Although no clear evidence of an overall power-law relationship was found for the magnetic hysteresis in a sinusoidally oscillating field in two dimensions~\cite{Thomas,Sides98,Sides99}, recently, with properly logarithmic corrections a dynamic scaling near a temperature other than the equilibrium transition point $T_0$ was found for the cooling FOPTs in the 2D Potts model~\cite{Pelissetto16}. This result shows that spinodal-like dynamic scaling does exist for FOPTs in systems with short-range interactions if logarithmic corrections are properly considered. However, in that case only one hysteresis exponent found numerically is consistent with a similar theory~\cite{Liang16}.

Here we first compare results arsing from both FT~\cite{Sides98,Sides99} and BD~\cite{Ryu,Ryu1} and numerical simulations of the 2D Ising model. We see that both the theories agree quite well generally with the numerical results. However, the slight but systematic deviations for different sweeping rates of the external driving indicate that the theories alone cannot explain such a driven transition. Then we find good agreement with the RG theory of FOPTs including instability exponents and even scaling forms as well as existence of finite instability points for two different $T$ below $T_c$ after account of additional logarithmic corrections. This offers compelling evidence for the theory and thus one can study the universality and scaling of FOPTs similar to their continuous counterpart.

\paragraph*{Finite-time scaling}
Crucial in our analysis is the theory of finite-time scaling (FTS)~\cite{Gong,Gong1}. We drive the FOPT by linearly rather than sinusoidally varying $H$. This linear driving is a direct implementation of the FTS~\cite{Gong,Gong1}, whose essence is a constant finite time scale associating with the given sweeping rate $R$ of the field. This single externally imposed time scale can thus probe effectively the transition when it is of the order of the nucleation time. In contrast, the sinusoidal driving has two controlling parameters, the field amplitude $H_0$ and the frequency $\omega$, and thus complicates and conceals the essence of the process~\cite{Feng}. In particular, at a fixed $H_0$, for $\omega\rightarrow 0$, the hysteresis loop area is governed by $H_0\omega$, which is equivalent to $R$, and increases with $\omega$; while for $\omega\rightarrow\infty$, the area is determined by $H_0^2/\omega$ in mean field and vanishes~\cite{Rao}. At least these two mechanisms compete and produce an area maximum at some $\omega$~\cite{Rao,Char,Sides98,Sides99}. In addition, for high $\omega$, the hysteresis loops are rounded and even not close and thus their areas are not well defined~\cite{Sides98}. This shortcoming does not contaminate the linear driving~\cite{Zhonge,Zhonge1}.

\paragraph*{Deficiency of nucleation theories for driving}
In FT~\cite{Sides98,Sides99}, if a positive constant $H$ is applied against $-M_{\rm eq}$, the field-theoretically corrected nucleation rate $I(T,H)$ per unit time and volume is given by~\cite{Langer67,Gunther}
\begin{equation}
I=B(T)H^Ke^{-F_c/k_{\rm B}T}= B(T)H^Ke^{-\Xi/H}\label{ith}
\end{equation}
with $\Xi=\Omega_2\sigma_0^2/2M_{\rm eq}k_{\rm B}T$ (see Supplemental material for details), where $F_c$ is the free-energy cost for the critical nucleus, $B(T)$ is a parameter, $K=3$ for the 2D kinetic Ising model~\cite{Langer67,Gunther,Rikvold94,Harris84}, $\Omega_d(T)$ is a shape factor in a $d$-dimensional space, $\sigma_0$ is the surface tension along a primitive lattice vector, and $k_{\rm B}$ is Boltzmann's constant.

In the MD regime, Avrami's growth law~\cite{Avrami} gives the magnetization $M$ at time $t$ as~\cite{Avrami,Sides99,Ramos}
\begin{equation}
M(t)=1-2\exp\left\{-\Omega_d\int_0^tI\left[\int_{t_n}^tv(t')dt'\right]^ddt_n\right\},\label{mt}
\end{equation}
where $v(t)$ is the interface velocity of a growing droplet. $v\approx g H^\theta$ with $\theta=1$ and a constant proportionality $g$ in the Lifshitz-Allen-Cahn approximation~\cite{Lifshitz,Lifshitz1,Gunton83}.

For a time-dependent field $H(t)=Rt$, by assuming an adiabatic approximation in which the constant field is simply replaced with its time dependent one~\cite{Sides99}, Eqs.~(\ref{ith}) and (\ref{mt}) then result in
$\Gamma(-4,x)/x^4-\Gamma(-6,x)/x^2+ \Gamma(-8,x)=4R^3\ln2/[\Omega_2g^2B(T)\Xi^8]$
with $x\equiv\Xi/H_c$ in two dimensions, where the coercivity $H_c$ is the field at $M=0$ and $\Gamma$ is the incomplete gamma function. An identical equation has been derived for the sinusoidal driving in the low frequency approximation~\cite{Sides99} in which
$R=H_0\omega\equiv2\pi H_0/[\tau(H_0,T)R_0]$ with $\tau(H_0,T)$ being the average lifetime of the metastable state at $H_0$ and $T$~\cite{Sides99}.

In the SD regime~\cite{Rikvold94}, by neglecting the growth time for a supercritical nucleus to occupy half the system volume $L^d$ compared with the nucleation time, the probability for the system to make the transition by time $t$ is~\cite{Sides98}
\begin{equation}
P(t)=1-\exp\left[-L^d\int_0^tI(T,H)dt\right].\label{pt}
\end{equation}
Accordingly, $H_c$ is approximately given by the time $t_c$ at which $P(t_c)=1/2$. Using again the adiabatic approximation for $I$, one obtains in this regime in two dimensions~\cite{Sides98} $\Gamma(-4,x)/x^4 =R\ln2/[B(T)L^2\Xi^4]\equiv CR$.

In BD, on the basis of the Becker-D\"{o}ring theory of nucleation~\cite{Becker2,Ryu,Ryu1}, the nucleation rate can also be cast in the form of Eq.~(\ref{ith}) but with a complicated $B(T,H)$ that
is $H$ dependent (see Supplemental material). $H_c$ in the MD and SD regimes can then be found similar to FT.

\begin{figure}[t]
\centerline{\includegraphics[width=1.0\linewidth]{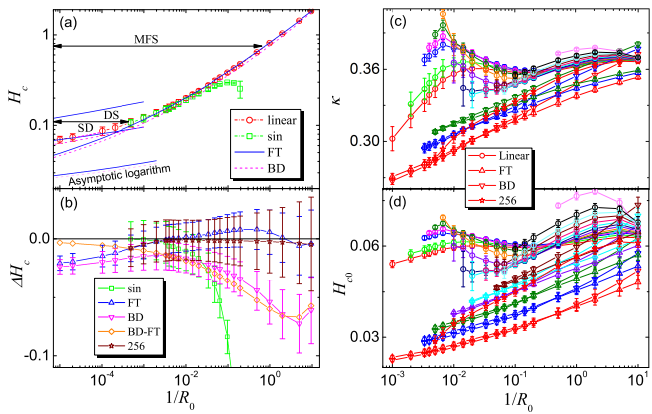}}
\caption{\label{hsr}(Color online) (a) $H_c$ versus scaled sweep rate $R_0$. Linear and sin indicate the data obtained numerically from the 2D Ising model using a linearly and a sinusoidally varying external field, respectively. Note that the ``error bars" give the standard deviations of the distributions of the transition involved~\cite{Sides99}. The three curves around SD are theoretical results for the single-droplet regime [one BD and two FT curves with $B(T)=0.02515$ for the upper and $B(T)=69.73$ for the lower] and the two lower curves are results of the asymptotic logarithmic approximation [the results of the larger $B(T)$ are far smaller and absent]. The horizontal lines with arrows indicate the dynamic spinodal (DS) and the mean-field spinodal (MFS)~\cite{Tomita,Rikvold94}. (b) Differences in $H_c$. BD-FT denotes the differences of the two theories, while the others are the differences to the linear driving. $256$ symbols the results about the $256^2$ lattices. (c) and (d) Finite time effects of $\kappa$ and $H_{c0}$, respectively. Each curve is obtained by successively omitting the datum with the smallest $R_0$ and plotting the results at the remaining smallest $R_0$. Different curves start with different largest $R_0$. The widths of the distributions have not been included into the fits, since their inclusion only slightly change the results of large $R_0$ for large ranges. For clarity, we plot only every other curve for the theories. Lines connecting the symbols are only a guide to the eye.}
\end{figure}

An asymptotic form $H_c\sim[-\ln (CR)]^{-1}$ can be found by expanding $\Gamma(a,x)$ in large $x$ in the SD regime~\cite{Sides98,Sides99}. This was argued to be the leading behavior for small $R$~\cite{Thomas}. However, it has been shown that such a behavior if exists could only be detected for extremely low $R$~\cite{Sides98,Sides99}, as seen by the curves marked asymptotic logarithm in Fig.~\ref{hsr}(a). We shall thus not pursue it.

Figure~\ref{hsr} shows the simulation results (see Supplemental material for detailed method) along with theoretical ones from solving numerically the relevant equations and their BD counterparts. Using the values of $H_c$ at $R_0=200$ in the linear driving, we find $B(T)=0.02515$, which is close to $0.02048$ found in Ref.~\cite{Sides99} but produces better results. As seen in Fig.~\ref{hsr}(a), the predictions of FT are excellent in the MD regime and even beyond, while in the SD regime, they are poor. To match the lowest rate, we find $B(T)=69.73$, larger by more than two thousand times. On the other hand, BD yields good results even remarkably in the SD regime without any adjustable parameters, though they are slightly smaller as seen in Fig.~\ref{hsr}(b) and the $H$ range is far larger than $0.01$ to $0.13$ studied in Refs.~\cite{Ryu,Ryu1} for a constant field.

Even though Fig.~\ref{hsr}(a) appears to demonstrate both FT and BD are quite good generally, comparing with other curves in Fig.~\ref{hsr}(b), one sees that both theories exhibit systematic deviations from the numerical results. This can be clearly seen from Figs.~\ref{hsr}(c) and (d), where we show the results of a systematic fits to the simple power law~\cite{Zhonge,Zhonge1},
$H_c=H_{c0}+aR_0^{-\kappa}$,
with constants $H_{c0}$, $a$, and $\kappa$. For the theories, both $\kappa$ and $H_{c0}$ change continuously with the range of $R_0$ that is used to find them, even if we change $\theta$ and $K$ to give better agreement with the numerical results, conforming to the expectation that the results described by such theories exhibit no scaling~\cite{Sides98,Sides99}. However, the simulation results are qualitatively distinct. If we include the theoretical data from the SD regime into the fits, we see a similar upturn near $R_0=10$ and a descent at larger $R_0$. This would indicate that the feature of the simulation results were related to a crossover from the MD to the SD regimes. However, deviations from the theoretical upturn are large (see Supplemental material for details). If we neglect in Figs.~\ref{hsr}(c) and (d) the two rightmost data, we see monotonic variations roughly up to the 12th curve (light cyan). This implies that the theories might be valid within the range from $R_0=0.5$ to $100$ or so, albeit not from the mean-field spinodal (MFS) above which spinodal decomposition occurs to the dynamic spinodal (DS) that separates regimes of MD and SD~\cite{Tomita,Rikvold94}.
However, Fig.~\ref{hsr}(d) shows clearly that there still exists a substantial discrepancy in $H_{c0}$ between the theories and the numerical results even in the reduced range, though $\kappa$ may agree. Note that this large gap cannot be removed by adjusting parameters like $B(T)$, because bigger $H_{c0}$ leads to bigger $\kappa$ and thus the gap transfers to $\kappa$. Moreover, such possible adjustments have only a
negligible effect since the differences in $H_c$ between the theories and the numerical results are small.

\begin{figure*}
\centerline{\includegraphics[width=0.8\linewidth]{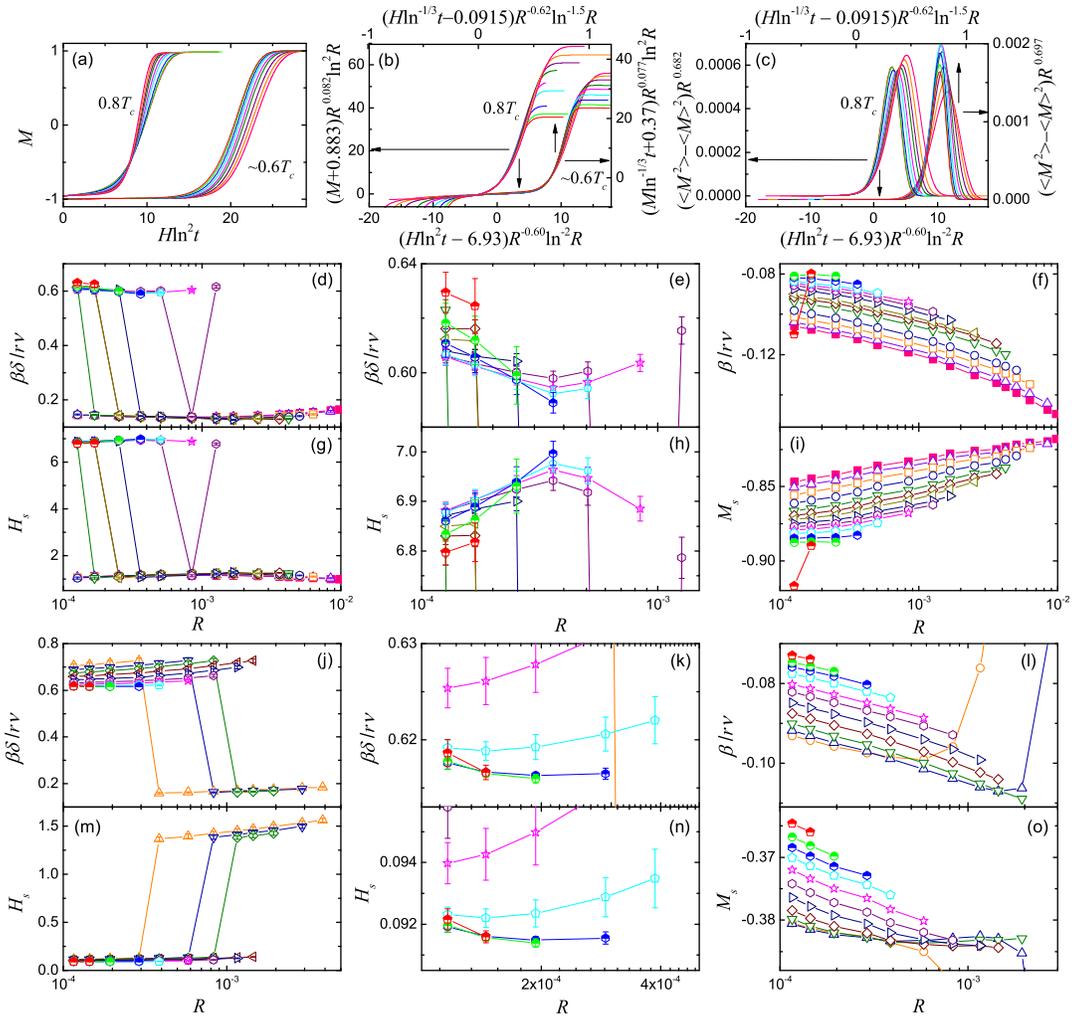}}
\caption{\label{rgt}(Color online) (a) $M$ versus $H\ln^2t$ for nine $R$ about from $0.0421$ to $0.000168$ (from right to left above the crossing) at $T=0.8T_c$ and from $0.00145$ to $0.000116$ (from left to right) at $T\approx0.6T_c$. (b) Rescaled of those curves in (a). (c) Rescaled of $\langle M^2\rangle-\langle M\rangle^2$. Note that only the rising parts of the curves are expected to collapse after being rescaled in line with (b). In (b) and (c), the arrows indicate the bottom-left and top-right axes used for the two $T$. (d) to (o) Finite time effects of $\beta\delta/r\nu$ and $H_s$ and $\beta/r\nu$ and $M_s$ fitted out of Eqs.~(\ref{hscaling}) and (\ref{mscaling}), respectively, for the four $n$ and $m$ given in (b). Different from Fig.~\ref{hsr}, for each curve, starting from the rightmost data point that represents the fit to the $R$ it stands and five others which are larger than it, each connected successive point denotes the fit of its $R$ and all the foregoing ones. The panels on the middle column zoom in on the corresponding panels on the left. (d) to (i) are results of $T=0.8T_c$ and (j) to (o) are the corresponding ones of $T\approx0.6T_c$. (d) and (g) [(j) and (m)] along with their respective enlarged ones (e) and (h) [(k) and (n)] are results of the fits of the $H$ at $M_s=-0.883$ [$-0.37$] for various $R$ at $T=0.8T_c$ [$T\approx0.6T_c$] and (f) and (i) [(l) and (o)] are those of the fits of the $M$ at $H_s=6.93$ [$0.0915$]. (g) and (h) [(m) and (n)] show that $H_s$ is roughly $6.93$ [$0.0915$] for the four curves (green, blue, cyan, and magenta) [the lowest three curves (red, green, and blue)], while (i) [(o)] shows that $M_s$ is about $-0.883$ [$-0.37$] for the corresponding data. They are thus self-consistent in that at $M_s$ the curves produce $H_s$ and just at this $H_s$ they give back to $M_s$ correctly. Errorbars are not shown in (f), (i), (l), and (o) since they are relatively large possibly due to the negative power of the logarithms, though the fits are good for a not-large $R$ range. Also the fits in these four panels appear not so approaching one another or level off as the others show, possibly because sub-leading contributions and corrections to scaling are stronger for $M$. Lines connecting symbols are only a guide to the eye. In (a) to (c), the data points are dense and only lines connecting them are displayed.}
\end{figure*}

\paragraph*{Evidence for the RG theory}
We next show that the $\phi^3$ theory can explain the results. Within the theory, scaling exists similar to the critical phenomena. For example, the scaling form for $M$ is~\cite{Zhongl05,zhong16},
$M\ln^{m}\!t=M_s+R^{\beta/{r\nu}}(-\ln R)^{m_1}f[(H\ln^{n}\!t-H_{s})R^{-\beta\delta/r\nu}(-\ln R)^{n_{1}}]$,
where $\beta$, $\delta$, $\nu$, and $r=z+\beta\delta/\nu$ are instability exponents for $M$, $H$, the correlation, and $R$, respectively, with $z$ being the dynamic exponent, each corresponding to its critical counterpart~\cite{Zhongl05,zhong16}, and $f$ is a scaling function. When $n=m=0$, $H_s$ and $M_s$ compose simply the instability point around which the theory is expanded and are thus finite, in sharp contrast with the critical phenomena. In the presence of the special logarithmic corrections in $t$, the point appears effectively at $H_s\ln^{-n}\!t$ and $M_s\ln^{-m}\!t$, which are scale dependence in consistent with previous studies~\cite{Binder78,Kawasaki,Kaski}. The $\ln^n\!t$ term with $n=d/(d-1)$ was argued to arise from the interplay between the exponential time in tunneling between the two phases and droplet formations in the low-$T$ phase in the Potts model~\cite{Pelissetto16}. In that case, the field is replaced by $T-T_0$. The curves of normalized energies versus $(T-T_0)\ln^2t$ for various cooling rates cross at a finite value, which was suggested to show a dynamic transition with spinodal-like singularity~\cite{Pelissetto16}. Figure~\ref{rgt}(a) shows that this crossing does appear for the Ising model studied here at $T=0.8T_c$. However, it is absent at $T\approx0.6T_c$. This indicates that the mechanism can not be dominated generally, as varying $T$ and varying $H$ cannot change the mechanism. We thus regard $n$ as an adjustable parameter and introduce generally the other exponents for the logarithmic corrections.

Our task is to show that the scaling form can indeed account for the data. This demands that there exist a single point, ($H_s,M_s$), such that at the particular $M_s$
\begin{equation}\label{hscaling}
H\ln^{n}\!t=H_{s}+a_1R^{\beta\delta/(r\nu)}(-\ln R)^{-n_{1}},
\end{equation}
while at the corresponding $H_s$,
\begin{equation}\label{mscaling}
M\ln^{m}\!t=M_{s}+f(0)R^{\beta/(r\nu)}(-\ln R)^{m_{1}},
\end{equation}
self-consistently, where $a_1$ is a constant satisfying $f(a_1)=0$. In order to reduce the parameters to be fitted and lift precision, we choose the values of the four $n$ and $m$ as input. We find this condition is highly restrictive for their choices. For example, if all $n$ and $m$ are set to zero, the condition cannot be satisfied. Neither can the seemingly plateau in Fig.~\ref{rgt}(o). In addition, since we have not considered sub-leading contributions and corrections to scaling, Eqs.~(\ref{hscaling}) and (\ref{mscaling}) are not expected to hold for a large range of $R$. Nevertheless, we require that the exponents obtained should somehow not depend on $R$ in a certain range.

Figure~\ref{rgt}(d) to (o) show the results. Except for (f) and (i), all other figures show that the fitted results exhibit jumps from large to small $R$ values. It is remarkable that when the self-consistent $H_s$ and $M_s$ are reached, the fitted results minimize their variations with $R$ and approach each other for some $R$ ranges. For example, at other $M_s$, the three lowest curves in Figs.~\ref{rgt}(k) and (n) tilt and separate from each other. For $T\approx0.6T_c$, $n=m=-1/3$ is not special. They can lie in the range between $-0.2$ to $-0.45$, with $\beta\delta/r\nu$ and $\beta/r\nu$ varying from $0.589$ to $0.635$ and from $-0.077$ to $-0.078$, respectively. The final fitted results are employed to collapse $M$ and its fluctuation $\langle M^2\rangle-\langle M\rangle^2$. The latter is rescaled just by $R^{(\beta\delta-\beta)/(r\nu)}$ rather than follows the susceptibility $\partial M/\partial H$, though the exponents for the two functions are identical. This arises from the violation of fluctuation-dissipation theorem in the nonequilibrium driving~\cite{Feng}. The collapses as displayed in Figs.~\ref{rgt}(b) and (c) are reasonably quite good, noting that only the leading behavior is considered, thus confirming the results. Note however that data collapses are sometimes deceptive. We show in Supplemental material an example in which the collapse appears perfect but unreasonable.

Besides the existence of the single finite $H_s$ and $M_s$, the most striking result is that the estimated exponents and their deviations from results of both $T$, $\beta\delta/r\nu\approx0.61(3)$ and $\beta/r\nu\approx-0.082(6)$, agree remarkably with their three and two loop results of $0.575$ and $-0.0905$, respectively, especially the negative value of $\beta$ in two dimensions~\cite{zhong16}. Moreover, although why the two $T$ data take on $n$ and $m$ values of opposite signs and their consequences have yet to be explored, a possible reason being the proximity of the high $T$ to $T_c$, the scaling functions appear to be universal up to a proper overall displacement and scaling as seen in Figs.~\ref{rgt}(b) and (c). These therefore provide a compelling evidence for the RG theory.

I thank Professor Per Arne Ridvold for his information and Shuai Yin, Baoquan Feng, Yantao Li, Guangyao Li, and Ning Liang for their useful discussions. This work was supported by National Natural Science Foundation of China (Grant Nos. 10625420 and 11575297).

\end{document}